\documentclass[twocolumn,showpacs,preprintnumbers,amsmath,amssymb]{revtex4-1}

\usepackage{epsfig}
\usepackage{dcolumn}
\usepackage{bm}

\begin{document}

\preprint{\today} 

\title{Measurement of the radial matrix elements for the $6s ^2S_{1/2} \rightarrow 7p ^2P_J$ transitions in cesium}

\author{Amy Damitz$^{1,2}$, George Toh$^{2,3}$, Eric Putney$^{1,4}$, Carol E. Tanner$^5$ and D. S. Elliott$^{1,2,3}$}

\affiliation{%
   $^1$Department of Physics and Astronomy, Purdue University, West Lafayette, Indiana 47907, USA\\
   $^2$Purdue Quantum Science and Engineering Institute, Purdue University, West Lafayette, Indiana 47907, USA\\
   $^3$School of Electrical and Computer Engineering, Purdue University, West Lafayette, Indiana 47907, USA\\
   $^4$Department of Physics and Astronomy, University of New Mexico, Albuquerque, New Mexico 87131 USA\\
   $^5$Department of Physics, University of Notre Dame, Notre Dame, Indiana 46556, USA
}

\date{\today}

\begin{abstract}
We report measurements of the electric dipole matrix elements of the $^{133}$Cs $\ $ $6s\,^2S_{1/2} \rightarrow 7p\,^2P_{1/2}$ and $6s\,^2S_{1/2} \rightarrow 7p\,^2P_{3/2}$ transitions. 
Each of these determinations is based on direct, precise comparisons of the absorption coefficients between two absorption lines.  For the $\langle 6s\,^2S_{1/2}||r|| 7p\,^2P_{3/2} \rangle$ matrix element, we measure the ratio of the absorption coefficient on this line with that of the D$_1$ transition, $6s\,^2S_{1/2} \rightarrow 6p\,^2P_{1/2}$.  The matrix element of the D$_1$ line has been determined with high precision previously by many groups.  For the $\langle 6s\,^2S_{1/2}||r||  7p\,^2P_{1/2}  \rangle$ matrix element, we measure the ratio of the absorption coefficient on this line with that of the $6s\,^2S_{1/2} \rightarrow 7p\,^2P_{3/2}$ transition. 
Our results for these matrix elements are $\langle 6s\,^2S_{1/2}||r|| 7p\,^2P_{3/2} \rangle = 0.57417 \: (57)~a_0$ and $\langle 6s\,^2S_{1/2}||r||  7p\,^2P_{1/2}  \rangle = 0.27810 \: (45)~a_0$. These measurements have implications for the interpretation of parity nonconservation in atoms.
\end{abstract}

\maketitle 

\section{Introduction}\label{sec:introduction}
Precise determinations of radial matrix elements of electric dipole (E1) transitions are essential for advancing the study of parity-nonconserving (PNC) weak-force-induced interactions in atoms. These matrix elements are important for testing calculations of the PNC transition amplitudes $\mathcal{E}_{\rm PNC}$~\cite{BlundellSJ92, KozlovPT01, PorsevBD10, RobertsDF2013}, as well as for determining the scalar and vector transition polarizabilities~\cite{BlundellSJ92, DzubaFS97, Derevianko00, SafronovaJD99, VasilyevSSB02}. For PNC studies based on the $6s~^2S_{1/2} \rightarrow 7s~^2S_{1/2}$ transition in cesium, for example, the most essential E1 matrix elements are $\langle ms~^2S_{1/2}  || r || np~^2P_J\rangle$, where $m, n = 6$ or 7, and $J = 1/2$ or $3/2$. Over the years, most of these quantities have been measured~\cite{YoungHSPTWL94,RafacT98,RafacTLB99,DereviankoP02a,AminiG03,BouloufaCD07,ZhangMWWXJ13,PattersonSEGBSK15,GregoireHHTC15,TannerLRSKBYK92,SellPEBSK11,antypas7p2013,BouchiatGP84,TohJGQSCWE18,BennettRW99,Borvak14,PhysRevA.99.032504} to a precision of 0.15\% or better.  The least precise moments, prior to the present work, were $\langle 6s~^2S_{1/2} || r ||  7p~^2P_{J}\rangle$.  Disagreement between three recent experimental results~\cite{VasilyevSSB02,antypas7p2013,Borvak14} at the $\sim1$\% level motivated us to re-examine these transitions. In this paper, we report new measurements of these matrix elements in $^{133}$Cs to a precision of $0.10\%$ and $0.16\%$, completing the required set of precise determinations of $E1$ dipole matrix elements between the lowest $ms~^2S_{1/2}$ and $np~^2P_J$ states.

\begin{figure} [b!]
	  \includegraphics[width=8.5cm]{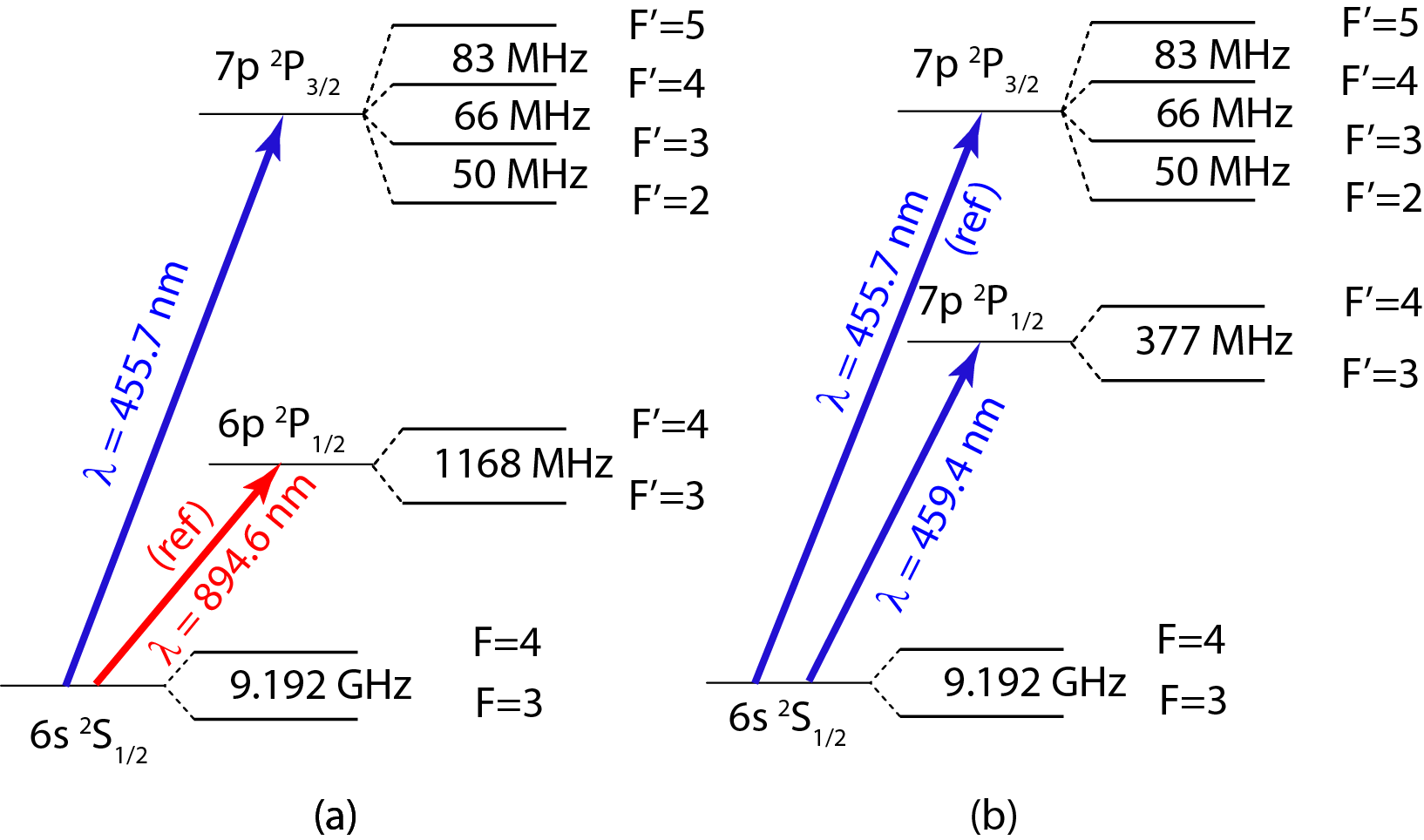}\\
	  \caption{Energy level diagrams of atomic cesium, showing the states relevant to these measurements. In (a), atoms are excited from the $6s\, ^2S_{1/2}$ ground state to the $6p\, ^2P_{1/2}$ ($\lambda =$ 894 nm) or the $7p\, ^2P_{3/2}$ ($\lambda =$ 456 nm) excited states, allowing for a comparison of the absorption coefficients of these two lines.
	  In (b), we compare the absorption coefficients for the $6s~^2S_{1/2} \rightarrow 7p~^2P_{1/2}$ line at 459 nm to that of the $6s~^2S_{1/2} \rightarrow 7p~^2P_{3/2}$ line at 456 nm.  Wavelengths shown in the figure are vacuum wavelengths.}
	  \label{fig:EnergyLevel}
\end{figure}

In order to determine the reduced matrix element for the $6s~^2S_{1/2} \rightarrow 7p~^2P_{3/2}$ transition at $\lambda = 455.7$ nm, we carry out a set of measurements in which we compare the absorption coefficient on this line to that of the `reference' D$_1$ line at 894.6 nm.  See the simplified energy level diagram in Fig.~\ref{fig:EnergyLevel}(a).
The matrix element for the latter is well measured~\cite{YoungHSPTWL94,RafacT98,RafacTLB99,DereviankoP02a,AminiG03,BouloufaCD07,ZhangMWWXJ13,PattersonSEGBSK15,GregoireHHTC15,TannerLRSKBYK92,SellPEBSK11}, with an impressive precision of 0.035\%. The ratio of absorption coefficients for these two lines therefore allows us to determine the reduced matrix element $\langle 6s~^2S_{1/2}  || r || 7p~^2P_{3/2}\rangle$ precisely. 
A similar comparison to the D$_1$ line strength for the $6s~^2S_{1/2} \rightarrow 7p~^2P_{1/2}$ transition at $\lambda = 459.4$ nm, however, is less fruitful. This is a weaker absorption line, and the difference between the absorption strength at 459 nm and at 894 nm is too great.  Therefore, we determine the matrix element at 459 nm through comparison with the 456 nm line strength, which now serves as the reference. We show the relevant transitions for this measurement in Fig.~\ref{fig:EnergyLevel}(b).

In each case, we use a pair of cw tunable single-mode diode lasers to measure and compare the absorption strengths of two lines in a cesium vapor cell. 
We direct the two laser beams through a cesium vapor cell along overlapping beam paths.  Then we block one laser beam to allow only the other to pass through the cell, scan the laser frequency through the resonance frequency, and record the absorption lineshape for this line. We then block the first laser, and record the absorption lineshape for the second.  We alternate measurements between lasers several times in quick succession.

These measurements differ from a previous measurement~\cite{antypas7p2013} from our group in several important regards.  In that measurement, we used a single blue diode laser which we could tune to either the $6s~^2S_{1/2} \rightarrow 7p~^2P_{1/2}$ transition at 459 nm or the $6s~^2S_{1/2} \rightarrow 7p~^2P_{3/2}$ transition at 456 nm, and compared the absorption coefficient of each of these lines directly to the absorption coefficient of the 894~nm line. The precision of the measurement for the 459~nm line suffered from the large difference in absorption strengths, as described previously.
In the present measurement, we avoid this difficulty by using two blue diode lasers, and determining the ratio of absorption coefficients for these two lines directly. Two additional improvements that we have made are: (1) In our previous measurement, we fit the absorption curves assuming a Gaussian Doppler-broadened lineshape.  We have discovered that, to attain a level of precision of less than 1\%, one must use a proper Voigt profile, a convolution of the homogeneous natural linewidth and the inhomogeneous Doppler-broadened linewidth, when the natural linewidths and/or Doppler widths of the two transitions differ from one another.  (2) For strongly absorbing lines, such as the D$_1$ line at the higher cell densities, the scan speed of the laser frequency becomes important. Under these strong absorption conditions, the medium changes quickly from fully transmitting to fully absorbing, and then back to fully transmitting again, as we tune the laser through the resonance.  If the rise and fall times of the photodetector are too slow, then one cannot obtain good fits to the data, and the measurement of the absorption coefficient is not reliable.  We have corrected each of these issues in the current measurements.

\section{Theory}\label{sec:theory}
 
When a low-intensity, narrow-band laser beam is incident upon a cell containing an absorbing atomic medium, the laser power transmitted through the cell can be written simply as 
\begin{equation}
P(\omega) = P_0 \exp{ \{-2 \alpha(\omega ) \ell_{cell} \} },
	  \label{eqn:fitfunction}
\end{equation}
where $P_0$ is the transmitted power in the absence of any absorption by the medium, $\alpha(\omega )$ is the frequency-dependent electric field attenuation coefficient, and $\ell_{cell}$ is the cell length.  The attenuation coefficient $\alpha(\omega )$ for linearly-polarized light by a Doppler-broadened atomic gas
in terms of the reduced $E1$ matrix elements $\langle J^{\prime} || \vec{r} || J \rangle $, as a sum over the various hyperfine components of the states, is given by Eq.~(14) of Ref.~\cite{RafacT98} as
\begin{eqnarray}\label{eq:abscoeff}
  \alpha(\omega) & = & \frac{2 \pi^2 n \alpha_{\rm fs} \omega }{\left( 2I+1\right) \left(2J+1\right) }
   \left| \langle J^{\prime} || \vec{r} || J \rangle \right|^2  \\ & & \hspace{0.3in}\times \sum_{F^{\prime}, } \sum_{F}  q_{J,F \rightarrow J^{\prime},F^{\prime}} V(\omega) , \nonumber
\end{eqnarray}
when the transition frequencies 
$\omega$ are independent of $m$ and $m^{\prime}$.  
$J$, $I$, and $F$ are quantum numbers for the total electronic, nuclear spin, and total angular momentum, respectively, and $m$ for the projection of $F$ on the $z$ axis. 
We use unprimed (primed) notation to indicate ground (excited) state quantities.
$n$ is the number density of the cesium atoms in the beam path, $\alpha_{\rm fs}$ is the fine structure constant, and
$q_{J,F \rightarrow J^{\prime},F^{\prime}}$ are weight  factors for the different hyperfine components due to the angular momentum of the states, 
\begin{eqnarray}\label{eq:q}
   q_{J,F \rightarrow J^{\prime},F^{\prime}} & = &(-1)^{2(I+J)} \left( 2F^{\prime} + 1\right) \left( 2F + 1\right)     \nonumber \\
  & & \hspace{-0.4in} \times \sum_{m, m^{\prime}}  \left( \begin{array}{ccc} F^{\prime} & 1 & F \\ -m^{\prime} & 0 & m \end{array} \right)^2 \left\{ \begin{array}{ccc} J^{\prime} & F^{\prime} & I \\ F & J & 1 \end{array} \right\}^2 .
\end{eqnarray} 
The arrays in the parentheses and curly brackets are the Wigner $3j$ (for linear polarization) and $6j$ symbols, respectively.  We list the values of $q_{J,F \rightarrow J^{\prime},F^{\prime}}$ for the transitions relevant to this work in Table \ref{table:q}.
\begin{table}[t!]
\caption{Numerical values of the weights $q_{J,F \rightarrow J^{\prime},F^{\prime}}$ for each of the hyperfine components of the $6s \ ^2S_{1/2} \rightarrow np \ ^2P_{j}$ transitions, as given in Eq.~(\ref{eq:q}). \label{table:q} }
  \begin{tabular}{|l|c|c|}
    \hline
      \multicolumn{1}{|c|} {$F \rightarrow F^{\prime}$}   		& $6s \ ^2S_{1/2} \rightarrow np \ ^2P_{1/2}$ & $6s \ ^2S_{1/2} \rightarrow np \ ^2P_{3/2}$  \\ \hline \hline
  
\rule{0in}{0.15in} $4 \rightarrow 3^{\prime} $ & $ 7/8$  & $7/48$ \\

\rule{0in}{0.15in} $4 \rightarrow 4^{\prime} $ & $ 5/8$ & $7/16$   \\

\rule{0in}{0.15in} $4 \rightarrow 5^{\prime} $ & $ - $	& $11/12$  \\

\rule{0in}{0.3in} $3 \rightarrow 2^{\prime}  $ & $ -$ & $5/12$   \\

\rule{0in}{0.15in} $3 \rightarrow 3^{\prime} $ & $ 7/24$ & $7/16 $ \\

\rule{0in}{0.15in} $3 \rightarrow 4^{\prime} $ & $ 7/8$ 	& $5/16$   \\
   \hline
  \end{tabular}
  
\end{table}

$V(\omega)$ is the Voigt lineshape function, 
\begin{equation}\label{eq:Voigt}
  V(\omega) = \sqrt{\frac{\ln 2}{\pi^3}} \frac{1}{\Delta \omega_D}  \int_{-\infty}^{\infty} \frac{\Gamma^{\prime} e^{-4 \ln2 \left( \omega_D / \Delta \omega_D \right)^2} d \omega_D }{\left[\omega  - \omega_D  - \omega_{F \rightarrow F^{\prime}} \right]^2 + \Gamma^{\prime 2}/4} ,
\end{equation} 
the convolution of the Lorentzian homogeneous lineshape function (of width $\Gamma^{\prime}$) and the Gaussian distribution of width $\Delta \omega_D$.
$\omega_{F \rightarrow F^{\prime}}$ is resonant frequency of the $F \rightarrow F^{\prime}$ hyperfine component of the transition.
This lineshape function is normalized such that its integral across the resonance is unity.  In this expression, $\omega_D$ is the Doppler shift, equal to $\omega v/c$, where $v$ is the atomic velocity, and $\Delta \omega_D$ is the Doppler full-width-at-half-maximum (FWHM) of the transition, equal to $\omega \: \sqrt{8 k_B T \ln2/(Mc^2)} \: $. 
Thus, precise measurements of the absorption in a cell would allow us to determine the radial matrix element for that individual transition, provided we also measure the vapor density in the cell, and the length of the cell.

Instead, by alternating transmission measurements between two spectral lines, we can determine the ratio of absorption strengths, and eliminate the need for precise determination of the cell length and vapor density.  
For example, to determine $\langle 6s_{1/2} || r ||  7p_{3/2}  \rangle$ (we abbreviate the state notation $|m \ell \ ^2L_J \rangle $ using only the quantum numbers of the single active electron $|m \ell_j \rangle $), we measure the absorption through the cell on the $6s \ ^2S_{1/2} \rightarrow 6p \ ^2P_{1/2}$ line at 894 nm, which serves as the reference, and the absorption on the $6s \ ^2S_{1/2} \rightarrow 7p \ ^2P_{3/2}$ line at 456 nm.  The ratio of matrix elements is then determined as
\begin{equation}\label{eq:ratioR1}
   R_1 \equiv \frac{\langle 6s_{1/2} || r ||   6p_{1/2}\rangle}{\langle 6s_{1/2} || r ||  7p_{3/2}\rangle} = \sqrt{\frac{\alpha_{3 \rightarrow 3^{\prime}}^{894}(\omega_0)/(7/24)}{\alpha_{F \rightarrow F^{\prime}}^{456}(\omega_0)/q_{F \rightarrow F^{\prime}}}}.
\end{equation}
$\alpha_{F \rightarrow F^{\prime}}^{\lambda}(\omega_0)$ is the attenuation coefficient at line center at wavelength $\lambda$ for the $F \rightarrow F^{\prime}$ component, and we have abbreviated the $q_{F \rightarrow F^{\prime}}$ factor, omitting the $J$ and $J^{\prime}$ for brevity from Eq.~(\ref{eq:q}).  
This factor $\alpha_{F \rightarrow F^{\prime}}^{\lambda}(\omega_0) / q_{F \rightarrow F^{\prime}}$ is the same for each of the different hyperfine components of the transition, so we define a scaled attenuation term
\begin{equation}
  \Upsilon^{\lambda} =\frac{ \alpha_{F \rightarrow F^{\prime}}^{\lambda}(\omega_0) \ell_{cell}}{ q_{F \rightarrow F^{\prime}}}
\end{equation}
for the line.  The term $\Upsilon^{\lambda}$ is the  attenuation of the absorbing vapor on the line at wavelength $\lambda$, 
defined in such a way as to make $\Upsilon^{\lambda}$ equivalent for each of the hyperfine components of the transition.
In terms of $\Upsilon$ then, the ratio $R_1$ is 
\begin{equation}\label{eq:ratioR1Upsilon}
   R_1 = \frac{\langle 6s_{1/2} || r ||  6p_{1/2}\rangle}{\langle 6s_{1/2} || r ||  7p_{3/2}\rangle} = \sqrt{\frac{\Upsilon^{894}}{\Upsilon^{456}}}.
\end{equation}
Similarly, we define the ratio 
\begin{equation}\label{eq:ratioR2}
   R_2 \equiv \frac{\langle 6s_{1/2} || r ||  7p_{3/2}\rangle}{\langle 6s_{1/2} || r ||  7p_{1/2}\rangle} = \sqrt{\frac{\Upsilon^{456}}{\Upsilon^{459}}},
\end{equation}
which we measure by comparing the attenuation coefficients of the $6s \ ^2S_{1/2} \rightarrow 7p \ ^2P_{3/2}$ line at 456 nm, which serves as the reference, and the absorption on the $6s \ ^2S_{1/2} \rightarrow 7p \ ^2P_{1/2}$ line at 459 nm. We describe our measurement of $R_2$ in Sec. \ref{sec:RME459}. 

There is an important subtlety regarding the role of the transition frequency $\omega$ on the attenuation coefficient.  The frequency $\omega$ appears in the numerator of the expression for the attenuation coefficient, Eq.~(\ref{eq:abscoeff}).  For a Doppler broadened medium, the Doppler width $\Delta \omega_D$, which is proportional to $\omega$, appears in the denominator of Eq.~(\ref{eq:Voigt}).  Therefore, for a Doppler-broadened transition, these frequency factors cancel, and the attenuation coefficients in Eq.~(\ref{eq:ratioR1Upsilon}) are independent of the optical frequencies of the two transitions.
Careful attention, however, must be paid to the proper normalization of the Voigt function.

\section{Measurement of $R_1$}\label{sec:RME456}
We first describe the measurement of the ratio of transition moments $R_1$, as defined in Eq.~(\ref{eq:ratioR1Upsilon}).  

\begin{figure}
	  \includegraphics[width=8cm]{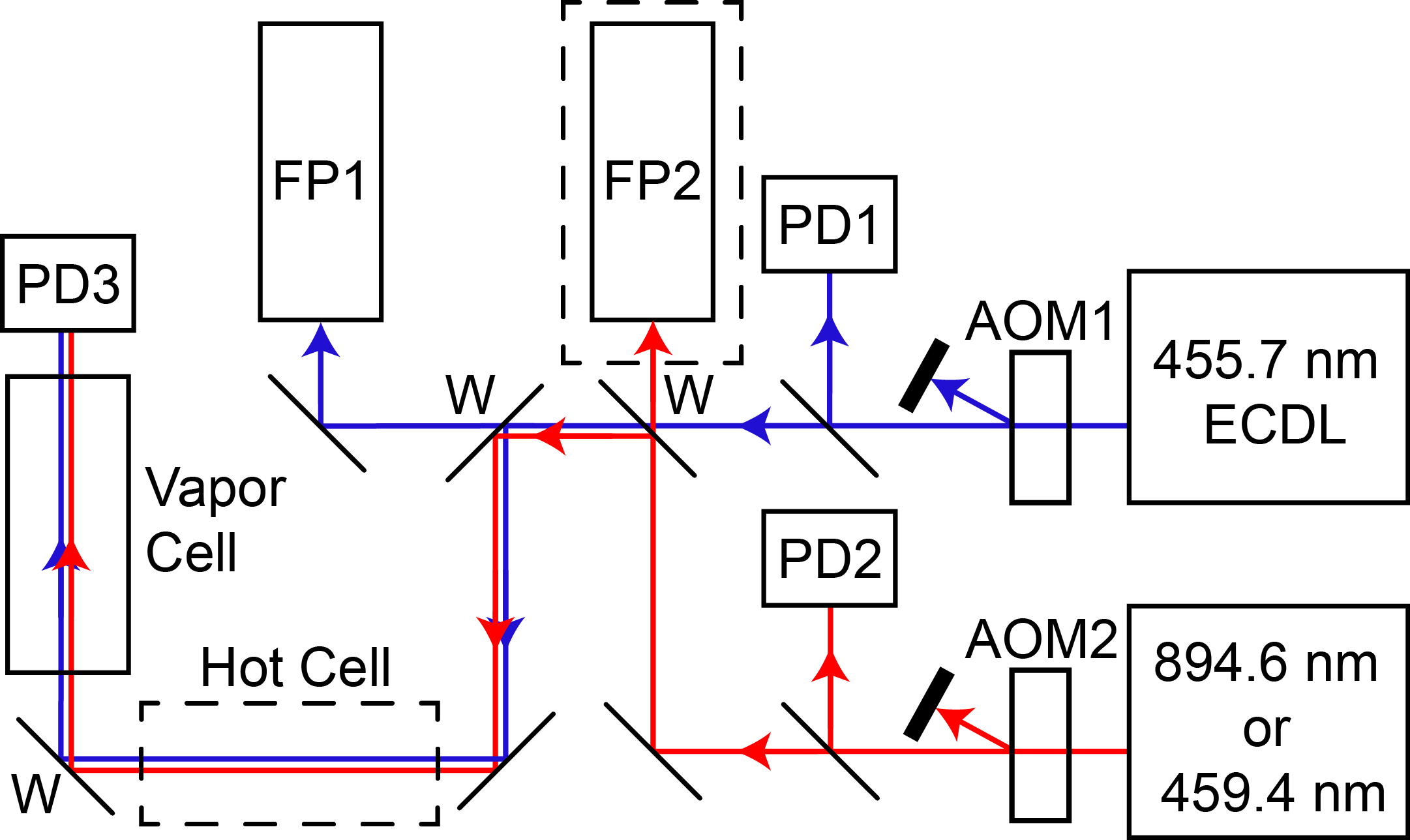}\\
	  \caption{Experimental setup for the two measurements. The 455.7 nm laser stays the same for both measurement. The 894.6 nm or 459.4 nm laser changes depending on the measurement, $R_1$ or $R_2$, respectively, being done. Abbreviations in this figure are: 
(AOM1,2) acousto-optic modulators; (ECDL) external cavity diode laser;
(PD1-3) photodiodes; (FP1,2) Fabry–-P\'erot cavities; and (W) wedged windows. FP2 (in the dashed box) is used only for the measurement of $R_1$.  }
	  \label{fig:ExperimentalSetup894}
\end{figure}

We show the experimental setup in Fig.~\ref{fig:ExperimentalSetup894}.
We use two home-made external cavity diode lasers (ECDL) in Littrow configurations, one at $\lambda = 894$ nm, the other at 456 nm.  The 894 nm laser produces $\sim$10 mW of output power, while the 456 nm laser produces $\sim$20 mW.  By ramping the laser diode current and the piezoelectric transducer (PZT) voltage concurrently, we are able to achieve mode-hop free scans of $7-10$ GHz, significantly greater than the widths of the spectra.

We align the laser beams so that they overlap one another in the cesium vapor cell, a sealed glass cell of inside length $\ell_{cell} = 29.9034 \: (44)$~cm fitted with flat windows. 
Control of the density of cesium in the cell is achieved using a cold finger enclosed within a copper block, whose temperature we control and stabilize to between $-8$ and $+18^\circ$C with a thermo-electric cooler and feed-back circuit.  We use Kapton heaters to keep the vapor cell above room temperature at $\sim 25^\circ$C, and enclose the cell in an aluminum shell inside an insulating styrofoam container to help maintain a stable and uniform body temperature. 
To detect the power of the laser beam transmitted through the vapor cell, we use a linear silicon photodiode, labeled PD3 in Fig.~\ref{fig:ExperimentalSetup894}. 
The photodiode current is amplified using a trans\-impedance amplifier with a gain of $5 \times 10^7$ V/A, designed for high-gain, low-noise operation.  This amplifier is followed by a second op amp with a gain of 10.  
We chose a slow scan rate ($\sim$4 GHz/s) and wide amplifier bandwidth (60 kHz), to allow fast rise- and fall-times of the signal.

\begin{figure*}[t!]
	\includegraphics[width=15cm]{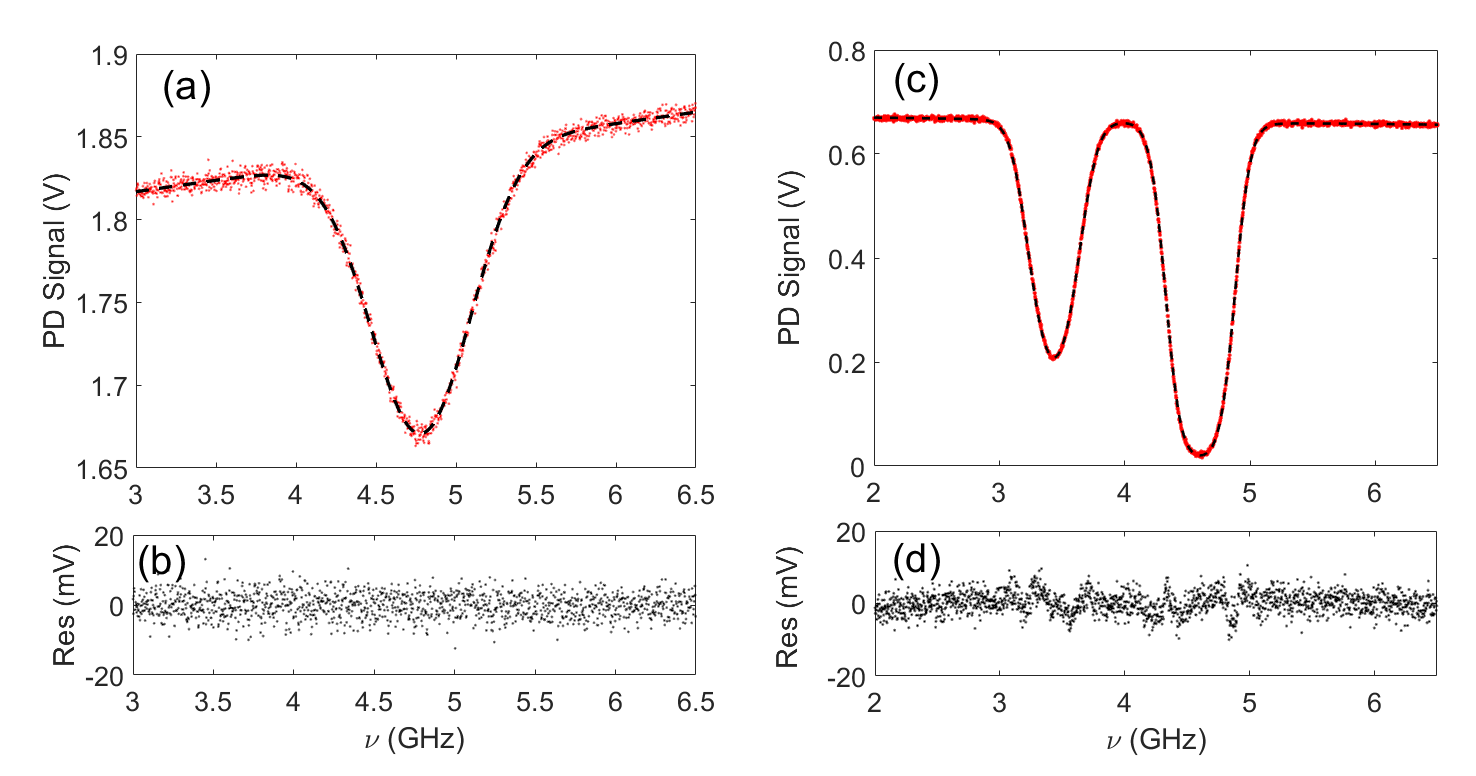}\\
	  \caption{Examples of absorption spectra recorded at a cold finger temperature of $-2^\circ$C.  These figures show the photodiode signal versus laser frequency as we scan (a) the 456 nm laser through $ F=4 \rightarrow  F'=3^{\prime},4^{\prime},5^{\prime}$ components (not resolved) of the $6s \ ^2S_{1/2} \rightarrow 7p \ ^2P_{3/2} $ transition, and (c) the 894~nm laser through $F=3 \rightarrow F'=3^{\prime},4^{\prime}$ components of the $6s \ ^2S_{1/2} \rightarrow 6p \ ^2P_{1/2} $ transition.  In each, the data are shown as the red data points, and the result of a least-squares fit as a black dashed line.  We show the residuals (data$-$fit) for each in (b) and (d).}
	  \label{fig:datafit894}
\end{figure*}

To improve the precision of the measurements, we stabilize the optical power delivered to the cell.  For this purpose, we diffract a fraction of each individual beam in an acousto-optic modulator (AOM1 or AOM2), and measure the relative power of the undiffracted beams using photodiodes (PD1 or PD2).  We use the photodiode current to generate an error signal, which controls the r.f.\ power applied to the AOMs.  In this manner, we are able to stabilize the power of each laser, and achieve a relatively flat power level for the scan of each laser, with less than 3\% variation in the laser power over a typical scan of $4-6$ GHz for the 456 nm laser and 0.5\% for the 894 nm laser.
To minimize saturation effects, the laser power incident on the cell from the 456 nm laser is about 40 nW in a $\sim$1 mm diameter beam and the 894 nm laser has 8 nW with $\sim$2 mm diameter. 
A 15 cm focal length lens after the vapor cell reduces the laser beam size incident on PD3 to less than the photocathode size.

We calibrate the frequency scans of the two lasers using separate Fabry-–P\'erot (FP) cavities, with free spectral ranges (FSR) of $\sim1500$ MHz. 
We record the transmission through the cavity concurrently with each absorption spectrum, and fit the frequencies of the transmission peaks to a 3rd-order polynomial in the laser frequency ramp voltage.

Before each set of measurements, we record the photodetector background offset voltage, the measured signal when no light is incident on the photodiode.
We also account for the small amount of laser power in the wings of the laser power spectrum. For this, we insert a second cesium vapor cell, which we heat to $\sim 120^\circ$C, into the beam path at the beginning of each data run. This vapor cell is labeled `Hot Cell' in Fig.~\ref{fig:ExperimentalSetup894}. Absorption in this cell of the on-resonant light is very strong, while off-resonant light is transmitted.
This gives us a good measurement of the laser power in the wings, typically $\sim 0.1\%$ of the total power for the 456 nm laser and $\sim 1\%$ for the 894 nm laser. 
We then determine the total offset level that comes from the background and laser power in the wings, to deduct from our data before curve fitting.

For each measurement, we block one of the lasers so that only one beam passes through the vapor cell, and record approximately four full absorption curves over a ten second period. We then block that laser, unblock the other, and record the absorption curves for the second laser. 
We repeat this process for a total of four records of 894 nm and three for 456 nm. In total, there are typically sixteen scans at 894~nm and twelve at 456~nm for each measurement.
Rapid reversals between the two wavelengths help minimize variations in the cesium density between measurements.
We perform multiple runs at each temperature. 
Then we change the temperature of the cold finger, wait for the cold finger temperature to stabilize, and collect new spectra. 
Additionally, we remove the absorption cell from the beam path and verify the absence of any spectral feature in the scans.

We show examples of absorption spectra at 456 nm and 894 nm in Fig.~\ref{fig:datafit894}.
The absorption peak at 456 nm, shown as the red data points in Fig.~\ref{fig:datafit894}(a), is made up of the three hyperfine transitions $F = 4 \rightarrow F^{\prime} = 3^{\prime}, 4^{\prime}$, and $5^{\prime}$. These peaks are unresolved since the hyperfine splitting of the $7p \ ^2P_{3/2}$ level~\cite{williamsHH18} is less than the Doppler width $\Delta \nu_D^{456} \sim 700$ MHz.  The slope of the unabsorbed signal to either side of the absorption dip is due to etalon effects (the variation of the transmitted power due to the interference between the reflections from the two window surfaces) in the windows of the cell. These windows are 1.2 mm thick, corresponding to a FSR of $\sim$80 GHz, far greater than the full 6 GHz scan length. 
The 894 nm absorption line shown in Fig.~\ref{fig:datafit894}(c) consists of two components, corresponding to $F^{\prime} = 3^{\prime}$ on the left and $F^{\prime} = 4^{\prime}$ on the right.  The frequency separation between these two peaks is the 
1167.7 MHz hyperfine splitting of the $6p \ ^2P_{1/2}$ state~\cite{UdemRHH99,Das2016,RafacTanner1997,GerginovCTMDBH06}, which is resolved in this spectrum since this splitting is greater than the $\Delta \nu_D^{894} \sim 360$ MHz Doppler broadening of the transition.  Note that the $F^{\prime} = 3^{\prime}$ peak is weaker than the $F^{\prime} = 4^{\prime}$, consistent with 
$q_{3 \rightarrow 3^{\prime}} = 7/24$ and $q_{3 \rightarrow 4^{\prime}} = 7/8$ for $J = J^{\prime} = 1/2$ 
listed in Table \ref{table:q}.  
For each of these lines, we have the choice of exciting from the $F = 3$ or $F = 4$ hyperfine component of the ground state.  Since the absorption strength of the 894 nm line is so much stronger than that of the 456 nm line, we used the $F = 3$ ground state (the weaker component) for the former, and $F = 4$ (the stronger component) for the latter. Given the limitations of our setup, the best case would be if both lines have the same strength, as we would be able to record data over a wider range of temperatures or attenuation coefficients.

For each of the spectra, we fit the data to an equation of the form shown in Eqs.~(\ref{eqn:fitfunction})$-$(\ref{eq:Voigt}). Our fit equation has five adjustable parameters: the level of full transmission, a term to account for the slope in the full-transmission level, the center frequency of one of the hyperfine components of the transition, and terms describing the Doppler-broadened width ($\Delta \nu_D^{\lambda}$) and amplitude ($\Upsilon^{\lambda}$) of the absorption dip.  The relative heights of the different hyperfine components are determined by the $q_{F \rightarrow F^{\prime}}$ factors in Table \ref{table:q}, and are fixed in our fits.  The linewidths in the Voigt lineshape are $\Gamma^{\prime} = 2 \pi (\Delta \nu_N  + 0.2 \ {\rm MHz}) $, where $\Delta \nu_N = $ 4.6 MHz~\cite{YoungHSPTWL94,RafacT98,RafacTLB99,DereviankoP02a,AminiG03,BouloufaCD07,ZhangMWWXJ13,PattersonSEGBSK15,GregoireHHTC15,TannerLRSKBYK92,SellPEBSK11} (1.22 MHz~\cite{PaceA75, MarekN76, GustavssonLS77, DeechLPS77, CampaniDG78, OrtizC81}) is the natural linewidth of the $6p ^2P_{1/2}$ line ($7p ^2P_{3/2}$ line), and 0.2 MHz is the intrinsic linewidth of the laser; and a Gaussian of width $\Delta \nu_D^{894} \sim 360$ MHz or $\Delta \nu_D^{456} \sim 700$ MHz. (We will return to this laser bandwidth correction later in this section.)  We use measured values for the frequency difference between the hyperfine components~\cite{RafacTanner1997,UdemRHH99,Das2016,williamsHH18}.  We show an example of the best fit profile to the absorption spectra as the black dashed lines in Figs.~\ref{fig:datafit894}(a) and (c).  In Figs.~\ref{fig:datafit894}(b) and (d), we show the residuals, the difference between our data and the fitted spectra. The small residuals indicate that the fits are very good models of the absorption profile. 

At the higher temperatures used for these measurements ($T > 10^{\circ}$C), we start to observe some departure of the ratio $\alpha_{3 \rightarrow 4^{\prime}}^{894}(\omega_0) / \alpha_{3 \rightarrow 3^{\prime}}^{894}(\omega_0)$ from the expected value of three when we fit the two peaks independently.  We suspect that this is a result of errors in the measurement of the offset voltage,  which become more critical for these strongly absorbing peaks.  At the most extreme temperature used ($T = 18^{\circ}$C), this ratio was as low as 2.946, so we attempted no measurements at higher temperatures. 

To include the effect of the spectral linewidth of the lasers in our fits to the absorption spectra, we first measured (1) the beat signal between the output of the 894 nm laser with that of a frequency comb laser (FCL); and (2) the beat signal between two similar blue diode lasers, each tuned to 456 nm.
In each case, we overlap the two interfering beams on a fast photodiode and observe the photocurrent on an r.f.\ spectrum analyzer. 
The long-term bandwidth in both cases was $2-3$~MHz.  In addition, 
we could observe the bandwidth of the signal on a single sweep of the spectrum analyzer. This shows considerable variation from sweep to sweep, probably due to acoustic vibrations of elements within the cavity, but we could observe lines as narrow as a few hundred kHz. We interpret these observations as a short-term (intrinsic) line width of $\Delta \nu_{Li} \sim200$~kHz, with slower fluctuations over a range of $\Delta \nu_{Ls} \sim3$~MHz. 
We calculate the effect of these laser frequency fluctuations, and determine that these can be included in the fits to the absorption spectra by modifying the Voigt lineshape function in two ways.  
First, we increase the homogeneous linewidth $\Delta \nu_N$, using the sum of the natural linewidth of the transition and the intrinsic linewidth $\Delta \nu_{Li}$ of the laser.  This is a small, but not negligible, increase.  
Second, we increase the inhomogeneous linewidth in the Voigt function calculation to the quadrature sum of the Doppler width, $\Delta \nu_D$, and the slow laser frequency fluctuations, $\Delta \nu_{Ls}$.  
For the linewidths of our system, this is a negligible increase.

\begin{figure}
  \includegraphics[width=\columnwidth]{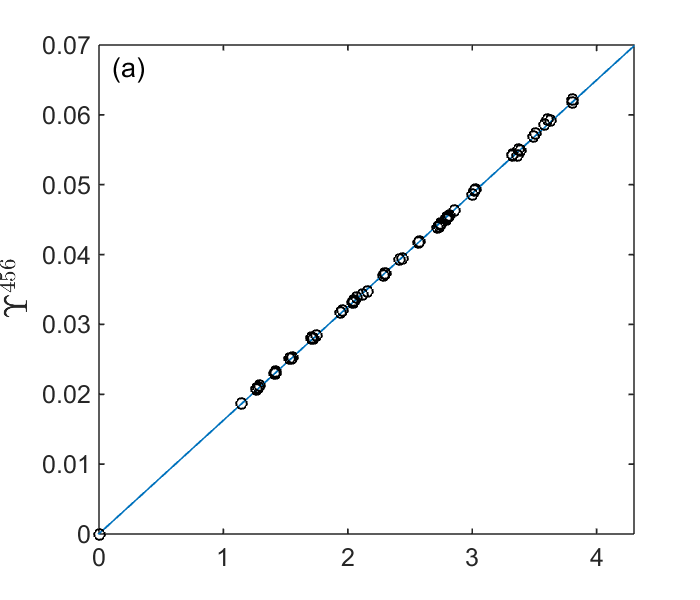}\\
  \includegraphics[width=\columnwidth]{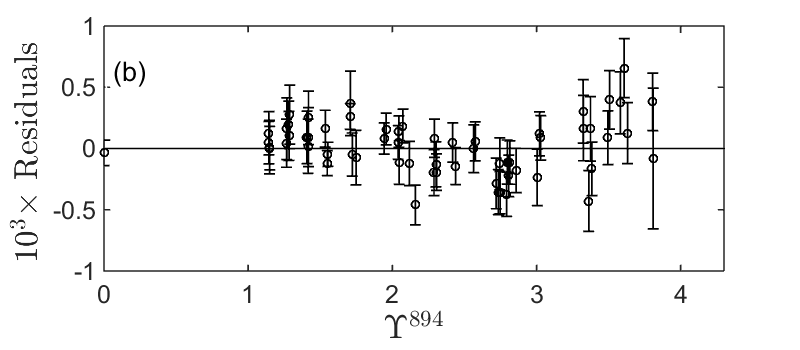}\\
	  \caption{
   Plot of $\Upsilon^{456}$ against $\Upsilon^{894}$ showing (a) the datapoints (circles) and the least squares fit straight line, and (b) the residuals in the ordinate. In (a), the error bars are smaller than the size of the markers. The vertical error bars in (b) represent the combined $1 \sigma$ uncertainties in $\Upsilon^{456}$ and $\Upsilon^{894}$. } 
	  \label{fig:fittedresult894}
\end{figure}

After fitting each of the sixteen (twelve) absorption curves at $\lambda = $ 894 nm ($\lambda = $ 456 nm) within a set individually, we compute the mean and standard deviation of the mean for the fitted values of $\Upsilon^{894}$ ($\Upsilon^{456}$). 
We show a plot of $\Upsilon^{456}$ vs.\ $\Upsilon^{894}$ in Fig.~\ref{fig:fittedresult894}. 
Each data point represents the average value of $\Upsilon^{456}$ and $\Upsilon^{894}$ at a particular cold finger temperature.  
The data point near the origin was recorded with the vapor cell removed from the beam path.  The error bars on the data points are too small to observe in Fig.~\ref{fig:fittedresult894}(a).  The vertical error bars in the residual plot, Fig.~\ref{fig:fittedresult894}(b), represent the combined $1 \sigma$ uncertainties in $\Upsilon^{456}$ and $\Upsilon^{894}$.

The total uncertainties in the values of $\Upsilon^{\lambda}$ are the statistical, etalon effects, and offset uncertainties added in quadrature. The statistical uncertainty comes from the standard deviation of the mean of the $\Upsilon^{\lambda}$ values from the fits to the sixteen (or twelve) absorption curves. 

The etalon effects are our estimate of the uncertainty of the attenuation coefficient resulting from the interference between the reflections at the cell windows.  We account for this effect to first order as a linear variation of the laser power with frequency, ignoring any curvature. 
This simple model is not adequate, however, when the peak or valley in the sinusoidal variation of the unabsorbed laser power is close to the frequency of the absorption feature.  In these frequency spaces, we estimate the effect of the curvature of the unabsorbed laser power on the size of the absorption peak, which we assign as the uncertainty due to the etalon effect.  In  cases of extreme curves, we also applied a small correction to the absorption height, along with an uncertainty of twice the size of the correction.  
We estimate the effect for each absorption curve depth.  A 1 mV change in the height of the signal changes $\Upsilon^{894}$ ($\Upsilon^{456}$) by 0.3\% (0.1\%).

The offset uncertainty is our estimation of the error in measuring the signal on PD3 resulting from the power in the wings of the laser spectrum and the background signal, which we subtract from the signal as an offset.  The offset uncertainty of the 456 nm signal is $\sim 0.05\%$ of $\Upsilon^{456}$, while for 894 nm it is more significant at $\sim 0.15\%$ due to the stronger absorption at 894 nm.

 \begin{figure*}
 	    \includegraphics[width=15cm]{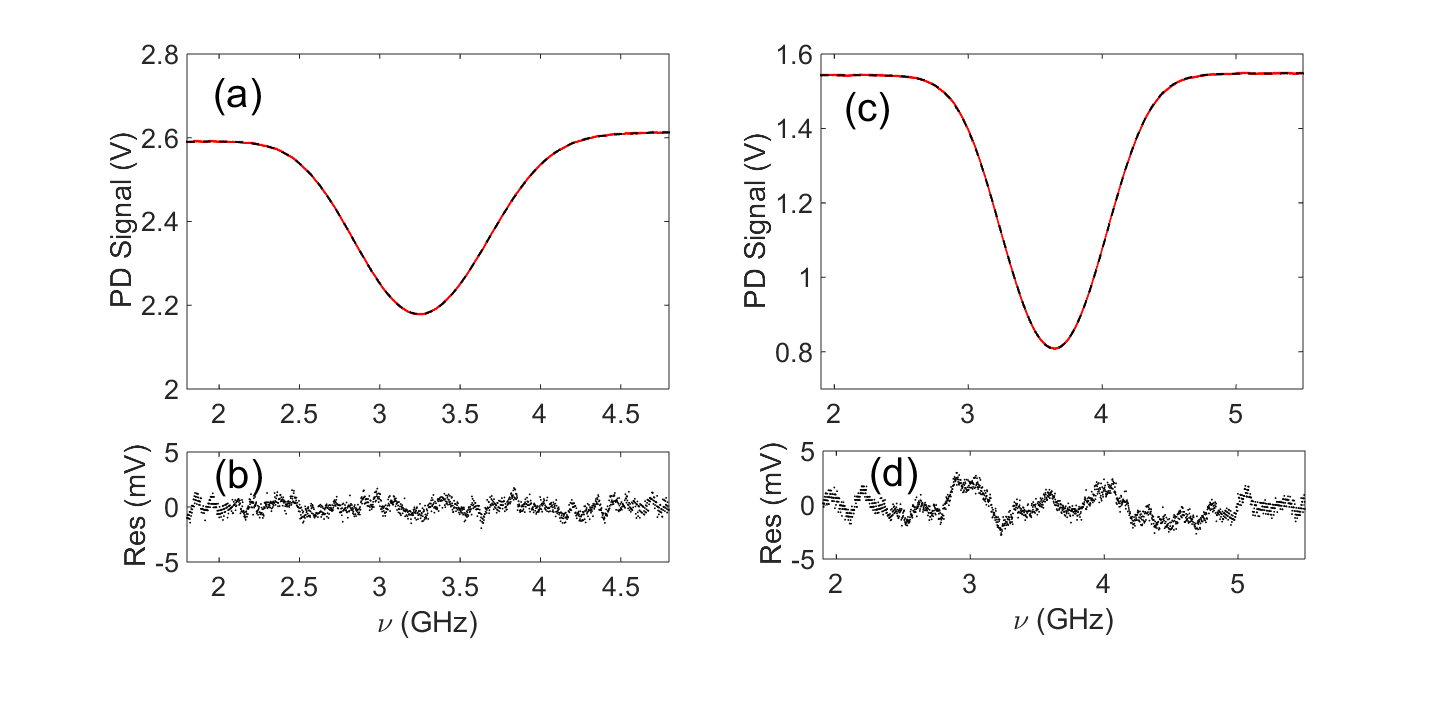}\\

 	  \caption{Examples of absorption spectra with the cell of length $\ell_{cell} \sim 6$ cm at a cold finger temperature of 56$^\circ$C.  These figures show the photodiode signal vs.~laser frequency as we scan (a) the 459 nm laser through $ F=4 \rightarrow  F'=3^{\prime},4^{\prime}$ components of the $6s \ ^2S_{1/2} \rightarrow 7p \ ^2P_{1/2} $ transition, and (c) the 456 nm laser through $F=4 \rightarrow F'=3^{\prime},4^{\prime},5^{\prime}$ components of the $6s \ ^2S_{1/2} \rightarrow 7p \ ^2P_{3/2} $ transition.  In each, the data are shown as the red data points, and the result of a least-squares fit as a black dashed line.  The hyperfine components are not resolved for either transition.  We show the residuals for each in (b) and (d). }
	  \label{fig:datafit1}
 \end{figure*}

The solid line in Fig.~\ref{fig:fittedresult894} is the result of a least-squares fit of a straight line to the data, with two adjustable parameters: the slope and the intercept.  We present the results of this fit in Table \ref{table:fits894}.  The intercept is within one standard deviation of zero, as expected, while the slope, i.e.~the ratio of scaled absorption terms, is $\Upsilon^{456} / \Upsilon^{894} = 0.016239 \: (21)$. (We show $1 \sigma$ uncertainties in the least significant digits inside the parentheses following the numerical value.)  The reduced $\chi_r^2$ for these data is 1.29, indicating that deviations of the data from the fitted line are slightly larger than the uncertainties would suggest. We expand the uncertainties of the slope by ~$\sqrt[]{1.29}$ to account for this, and quote a final slope of $0.016239 \: (23)$.
Using Eq.~(\ref{eq:ratioR1Upsilon}), the inverse of the square root of the slope yields the ratio of matrix elements $R_1 = 7.8474 \: (56)$. 
\begin{table}[t!]
    \caption{Numerical values for the intercept, slope, and reduced $\chi_r^2$ from the fit to the data in Fig.~\ref{fig:fittedresult894}. These uncertainties have not yet been expanded by $\sqrt[]{\chi_r^2}$. }
    \begin{tabular}{c|c}
    \hline
    Parameter \rule{0.1in}{0in} & Value  \\ \hline \hline
    \rule{0in}{0.2in}   Intercept  & \rule{0.1in}{0in}$3.6\ (43) \times 10^{-5}$  \\
    \rule{0in}{0.2in} Slope & $0.016239\ (21)$   \\
    \rule{0in}{0.2in}    $\chi_r^2$ & $1.29 $	  \\
    \hline
  \end{tabular}
  \label{table:fits894} 
\end{table}

In Sections \ref{sec:error} and \ref{sec:Results}, we will consider a few additional systematic effects that contribute to this measurement, and use $R_1$ to determine the E1 matrix element for the $6s \ ^2S_{1/2} \rightarrow 7p \ ^2P_{3/2}$ transition.  Before we do this, we describe parallel measurements of the absorption strength of the  $6s \ ^2S_{1/2} \rightarrow 7p \ ^2P_{1/2}$ transition.

\section{Measurement of $R_2$}\label{sec:RME459}

We turn now to the measurement of the ratio $R_2$, as defined in Eq.~(\ref{eq:ratioR2}).  
We carry out this measurement in a fashion similar to that of the measurement of $R_1$ discussed in Sec.~\ref{sec:RME456}, alternating between absorption measurements on 
the $6s \ ^2S_{1/2} \rightarrow 7p \ ^2P_{3/2}$ line at 456 nm and the $6s \ ^2S_{1/2} \rightarrow 7p \ ^2P_{1/2}$ line at 459 nm.
The experimental setup is very similar to the one discussed in the previous section, and is shown in Fig.~\ref{fig:ExperimentalSetup894}. The 894 nm ECDL is replaced with a 459 nm ECDL, and FP2 is removed, since we can use the same Fabry-P\'{e}rot cavity FP1 for both lasers.  
Each laser generates approximately 20 mW of laser light, and produces mode-hop free scans of $>7$ GHz.

Other differences in the apparatus or procedure include: ($i$) We carry out these measurements in three separate data sets, which differ in F, the hyperfine level of the ground state, or the vapor cell used.  This is in contrast to our determination of $R_1$, for which we use only one F value and one vapor cell. The first two data sets are performed with a 
short (of length $\ell_{cell} \sim 6$ cm) sealed glass cell mounted with $0.5^{\circ}$ wedged windows. The shorter cell length requires higher Cs densities for comparable absorption, and the wedged windows reduce the magnitude of the etalon effects.  We control the density of cesium in the cell using a cold finger enclosed within an aluminum block, whose temperature we control and stabilize to between $40-65^\circ$C using a thermo-electric module and feed-back circuit.  We use heat tape coiled around the vapor cell to heat the cell body to $\sim 80^\circ$C, 
and wrap the cell and heat tape with aluminum foil to help maintain a stable and uniform body temperature. 
For the third data set on these lines, we used the long ($\ell_{cell} = 29.9$ cm) vapor cell described in Sec.~\ref{sec:RME456}.
Observing similar results in this second cell allows us to rule out background gas in the cell or collisional effects as possible sources of error.
($ii$) Additionally, the linear silicon photodiode, labeled PD3 in Fig.~\ref{fig:ExperimentalSetup894}, had a slower rise/fall time, as both of the absorption curves were similar in depth and we could achieve a better signal-to-noise ratio by filtering out the high-frequency noise. 
We amplify the photodiode current using a transimpedance amplifier of gain $5 \times 10^7$ V/A and bandwidth of 2 kHz, followed by a second amplifier of gain 10, and measure $\sim 1$ mV of noise on a $\sim 2$ V signal. The laser power incident on the cell is about 50 nW in a $\sim$1 mm diameter beam for both lasers. 
($iii$) Finally, since the curves were shallower, we were able to scan the laser frequencies through the absorption curves more rapidly. When we investigated the effects of the bandwidth and scan rate as we did for $R_1$, we found that recording eight full absorption curves over a two second period allows good fits to the data. 

For these data, measurements of the background offset voltage several times each day, rather than before each run, were sufficient.  
The background offset voltage was small ($\sim 1$ mV), and variations were minimal, falling well within the measurement uncertainty. 
Before every run we did insert the hot cell to estimate and correct for the small amount of laser power in the wings of the laser power spectrum.
This gives us a good measurement of the laser power in the wings, 
typically $\sim 0.1\%$ of the full power incident on the photodiode for the 456 nm and $\sim 0.3\%$ for the 459 nm laser. We deduct the total offset in the signal, the power in the laser wings along with the background signal, from the data before fitting. We 
estimate the uncertainty of the attenuation coefficients due to the offset to be 0.05\% (0.1\%) for the 459 nm (456 nm) laser.

We show examples of the measured spectra as the red data points in Fig.~\ref{fig:datafit1}(a) ($6s \ ^2S_{1/2} \rightarrow 7p \ ^2P_{1/2}$ line at 459 nm) and \ref{fig:datafit1}(c) ($6s \ ^2S_{1/2} \rightarrow 7p \ ^2P_{3/2}$ line at 456 nm).  We fit the data to an equation of the form shown in Eqs.~(\ref{eqn:fitfunction})$- $(\ref{eq:Voigt}), using the same five adjustable parameters as described in Section \ref{sec:RME456}.
The lineshape of each hyperfine component of the transition is a Voigt profile, with a Lorentzian width $\Delta \nu_N$ (1.22 MHz for the $7p ^2P_{3/2}$ line, or 1.06 MHz for the $7p ^2P_{1/2}$ line~\cite{PaceA75, MarekN76, GustavssonLS77, DeechLPS77, CampaniDG78, OrtizC81}) added to the linewidth of the lasers of 0.2 MHz and a Gaussian of width $\Delta \nu_D^{456} \sim \Delta \nu_D^{459} \sim 750$ MHz.  We use calculated values for the relative amplitudes ($q_{F \rightarrow F^{\prime}}$ of Table \ref{table:q}) and experimental values~\citep{williamsHH18} for the frequency difference of the hyperfine components. We show the least-squares fit spectral profiles as the black dashed lines in Figs.~\ref{fig:datafit1}(a) and (c).  The residuals, the difference between the data and the fitted profile, are shown in Figs.~\ref{fig:datafit1}(b) and (d).
We fit each of the twenty-four (thirty-two) scans at 456 nm (459 nm) within a measurement individually, and compute the mean and standard deviation of the mean of the fitted values of $\Upsilon^{\lambda}$.

\begin{figure}
          \includegraphics[width=\columnwidth]{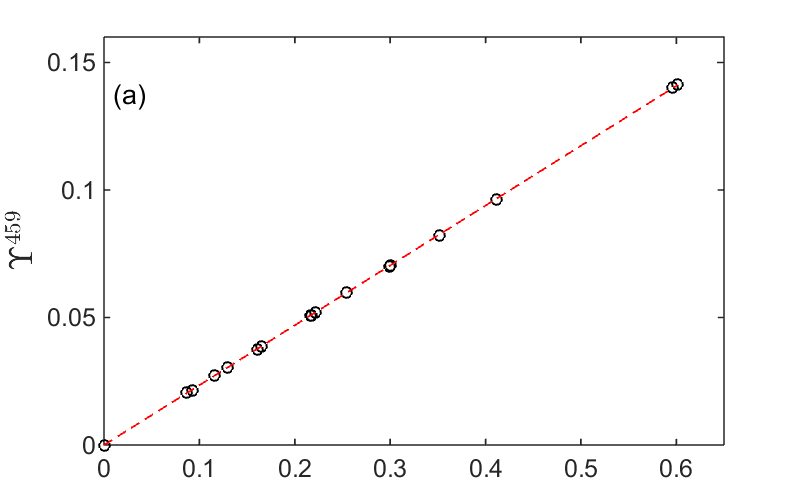}\\
          \includegraphics[width=\columnwidth]{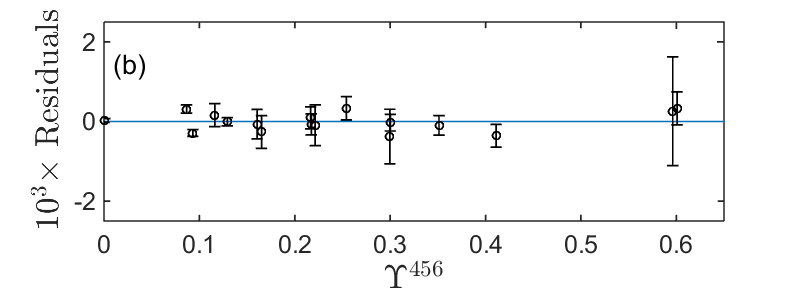}\\      
    	  \caption{Plot of $\Upsilon^{459}$ vs.\ $\Upsilon^{456}$ for data set 2 showing (a) the datapoints (circles) and line of best fit, (b) residuals in the ordinate. In (a), the error bars are smaller than the size of the markers. The vertical error bars in (b) represent the combined $1 \sigma$ uncertainties in $\Upsilon^{459}$ and $\Upsilon^{456}$.}
    	  \label{fig:fittedresult}
    \end{figure}

Finally, we plot $\Upsilon^{459}$ against $\Upsilon^{456}$, and determine the least-squares fit of a straight line to determine the slope.  An example of one such plot (set 2) for the transition from $6s \: ^2S_{1/2}, \ F=4$ is shown in Fig.~\ref{fig:fittedresult}(a). 
Each point on the graph corresponds to a different cold finger temperature, with the $y$-coordinate and $x$-coordinate coming respectively from the 459 nm and 456 nm average. 
We determined the uncertainties of each $\Upsilon^{\lambda}$ as the quadrature sum of
the statistical, etalon effect, and offset uncertainties. The etalon effect uncertainty is as described for $R_1$. 
We found that a 1 mV change in the offset resulted in a change to $\Upsilon^{456}$ and $\Upsilon^{459}$ of 0.1\% and 0.05\%, respectively. 
The error bars shown on the residual plot, Fig.~\ref{fig:fittedresult}(b), represents the combined errors of $\Upsilon^{459}$ and $\Upsilon^{456}$.
We perform separate analyses of the $6s \: ^2S_{1/2}, \  F=3$ and $F=4$ data, since their sensitivity to systematic effects differs.
We 
present the intercepts and slopes from individual straight line fits for the three data sets in Table \ref{table:2dayresults}.  The intercepts are again all acceptably close to zero.  
We will derive the ratio $R_2$ from the square root of the inverse of the slopes of these plots, as shown in Eq.~(\ref{eq:ratioR2}), but first we consider some additional systematic effects, as discussed in the following section.

 \begin{table}[t]
    \caption{Summary of linear fit results from our three data sets of $\Upsilon^{459}$ against $\Upsilon^{456}$. The listed uncertainties for the slope and intercept have not yet been expanded by $\sqrt[]{\chi_r^2}$.  Data sets 1-2 were collected with the shorter $\sim 6$ cm cell. Data for set 3 was recorded using the longer $\sim 30$  cm cell.}
   \begin{tabular}{|c|c|c|c|c|}
    \hline
 	   Data Set & Intercept \rule{0in}{0.15in}& Slope & $\chi_{r}^2$    \\ 
       \hline \hline 
       
   \rule{0.09in}{0in}Set 1, F=3\rule{0.09in}{0in}	\rule{0in}{0.15in}	& $-5.8\: (41) \times 10^{-5}$ &\rule{0.09in}{0in} $0.23380 \: (29) $ \rule{0.09in}{0in}& \rule{0.09in}{0in}$1.95$\rule{0.09in}{0in} \\
    Set 2, F=4 \rule{0in}{0.15in}  	& $-2.4 \: (42) \times 10^{-5} $ & $0.23477 \: (32)$  & $1.95$  \\

    Set 3, F=3  \rule{0in}{0.15in}		& $-1.4 \: (51)  \times 10^{-5}$ & $0.23593 \: (48)$  & $1.52$  \\

    \hline

   \end{tabular}

   \label{table:2dayresults}
 \end{table}

\section{Errors}\label{sec:error}
We have investigated several potential sources of systematic effects, listed in Table \ref{table:sourcesoferror}, to estimate their impact on the measurements. We describe each of these effects in this section. All of these systematic effects are applied to $R_1$ ($R_2$), after fitting $\Upsilon^{456}$ ($\Upsilon^{459}$) against $\Upsilon^{894}$ ($\Upsilon^{456}$).  

We derive the uncertainties labeled `Fit' from the fitted values of the slope and their uncertainties, listed in Tables \ref{table:fits894} and \ref{table:2dayresults}.  We have expanded these uncertainties by $\sqrt{\chi_r^2}$ to account for excess variation of the data points. 
\begin{table}[b!]
    \caption{Sources of error and the percentage uncertainty resulting from each, for the determinations of $R_1$ and $R_2$.   
    We add the errors in quadrature to obtain the total uncertainty. 
    }
    \begin{tabular}{|l|  c|  c|}
        \hline

      \rule{0.1in}{0in}Source             &   \rule{0.1in}{0in}$\sigma_1/R_1$(\%)	\rule{0.1in}{0in}	 & \rule{0.1in}{0in} $\sigma_2/R_2$(\%)\rule{0.1in}{0in} \\
		\hline \hline
        Fit					& 0.07  &0.09-0.13 \\
		
        Freq.~scan calibr.			& 0.04  & 0.01 \\
        Zeeman				&0.03 	&0.02 \\
        Beam overlap					& 0.01  & 0.01 \\

        Saturation		& 0.02 	& 0.02 \\
				
        Linewidth		& 0.02 	& 0.02 \\
                                       			
                                        \hline \hline
        Total uncertainty				& 0.09	&0.09-0.13\\
    \hline
    \end{tabular}

    \label{table:sourcesoferror}
\end{table}

 \begin{table}
  \caption{Summary of the three data sets of $R_2$ values after correction for the effects of the Zeeman splitting.  The uncertainty in $R_2$ includes the errors due to the fit (increased by $\sqrt{\chi_r^2}$ of the appropriate data set), frequency calibration, Zeeman splitting, beam overlap, saturation, and linewidth.  Data sets 1-2 were collected with the shorter ($\sim 6$ cm) cell. Data for set 3 was recorded using the longer ($\sim 30$ cm) cell. The $\chi_r^2$ of the weighted mean of the three sets is 4.2 and the weighted mean's error has been expanded by $\sqrt[]{4.2}$. See Fig.~\ref{fig:R2} for a plot of the results of the individual data sets and the weighted mean. }
   \begin{tabular}{|c|c|c|c|c|}
    \hline
 	   Data Set\rule{0in}{0.15in} & $R_2$  \\ 
       \hline  \hline
    Set 1, F=3\rule{0in}{0.15in} & 2.0684 (19) \\
    Set 2, F=4\rule{0in}{0.15in} & 2.0637 (21)\\
    Set 3, F=3\rule{0in}{0.15in} & 2.0591 (27) \\
    \hline
 \rule{0.1in}{0in}  Weighted Mean\rule{0in}{0.15in}	\rule{0.1in}{0in}	& \rule{0.1in}{0in}2.0646 (26)\rule{0.1in}{0in}\\ 
    \hline
   \end{tabular}
  
   \label{table:R2results}
 \end{table}

During the course of analyzing the absorption curves, we noted a sensitivity of the fits to the frequency calibration of the laser scans. We calibrate these scans, as discussed earlier, using the transmission peaks of the lasers through the FP cavities.  The FSRs of these cavities, however, are not known precisely. 
We experimentally determined the FSR values of both FP cavities by fitting the absorption data using different values of the FSR. Using the variation in the residuals, we obtain an estimate for the FSR that fits the absorption curves the best. 
(Since the hyperfine splittings of each of these states are well known, the residuals of the absorption curves are sensitive to variations in the frequency calibration of the scans.)
We determined the FSR for the FP cavity used with the 456/459~nm lasers to be $FSR_{FP1} = 1501.6 \: (10)$ MHz, while for the 894~nm laser, we determine $FSR_{FP2} = 1481.9 \: (4)$ MHz.
We also use these fits to estimate the effect that the uncertainty of the FP FSR has on the measured ratios. 
We estimate an uncertainty in the ratio $R_1$ due to frequency calibration, to be $0.04\%$. For the ratio $R_2$, we find the uncertainty to be at most $0.01\%$. 

The magnetic field at the cell location also affects the measurements of the absorption strength.  We measure a static magnetic field of $\sim 1$ G in the vertical direction (parallel to the laser polarization) at the location of the 6 cm vapor cell, mainly from the optical table. We minimize the magnetic field generated by the heat tape by wrapping the heat tape in alternating directions, ensuring the magnetic field only comes from the surroundings.  A  $\sim 1$~G field causes a Zeeman splitting on each hyperfine component of 2 MHz or less. 
We approximate the effect of Zeeman splitting on the effective homogeneous linewidth of the transition by adding the Zeeman broadening of each hyperfine component to the natural linewidth, which we use in the Voigt function for our analysis. 
We multiply $R_2$ by the Zeeman correction for the appropriate starting F state, 0.9999 for F=4 data and 1.0001 for F=3 data. We estimate an uncertainty in  $R_2$ due to this correction to be about $0.02 \% $.
The height of the 30 cm cell above the table was greater than that of the small cell, so the magnetic field for measurements of $R_1$ was smaller, $\sim$0.5 G. This led to a Zeeman splitting of less than 0.7 MHz.  We estimate the uncertainty in $R_1$ due to this Zeeman splitting to be 0.03\%, and did not apply any correction to these data.

Smaller systematic errors in the ratios result from beam overlap errors and saturation effects. We estimate that each of these effects contribute 0.02\% uncertainty or smaller, as listed in Table~\ref{table:sourcesoferror}.  
We measure that the two laser beams are parallel to one another to within 0.05 mrad, and overlap each other in the cell to within 0.5 mm.  Therefore the effective path lengths for these two beams are identical to within 0.02\%, for an effect on $R_1$ and $R_2$ of 0.01\%. 
We minimize saturation effects by reducing the laser intensity of the 456 nm and 459 nm lasers to less than $2~\times~10^{-4} $ times the saturation intensity for the transition~\cite{Urvoy11} using a neutral density filter and reflections from several uncoated wedged windows. We attenuated the power of the 894 nm laser more than the 456 nm and 459 nm laser to similarly avoid saturation.
We estimate that saturation effects could have an effect at the $0.02\%$ level.

Lastly, we include an uncertainty for the correction that we apply for the linewidth of the lasers used. For all of the ECDL lasers we estimated a 200 kHz intrinsic linewidth with a conservative uncertainty of 200 kHz. These uncertainties would lead to about a 0.02\% uncertainty in each of the ratios.

We add the fit, frequency scan calibration, Zeeman, beam overlap, saturation, and linewidth errors in quadrature for our final uncertainties to $R_1$ and $R_2$,
and apply the Zeeman correction to $R_2$ to get our final values. 
In the next section, we discuss the results of these measurements.

\section{Results}\label{sec:Results}

\begin{figure}
    \includegraphics[width=\columnwidth]{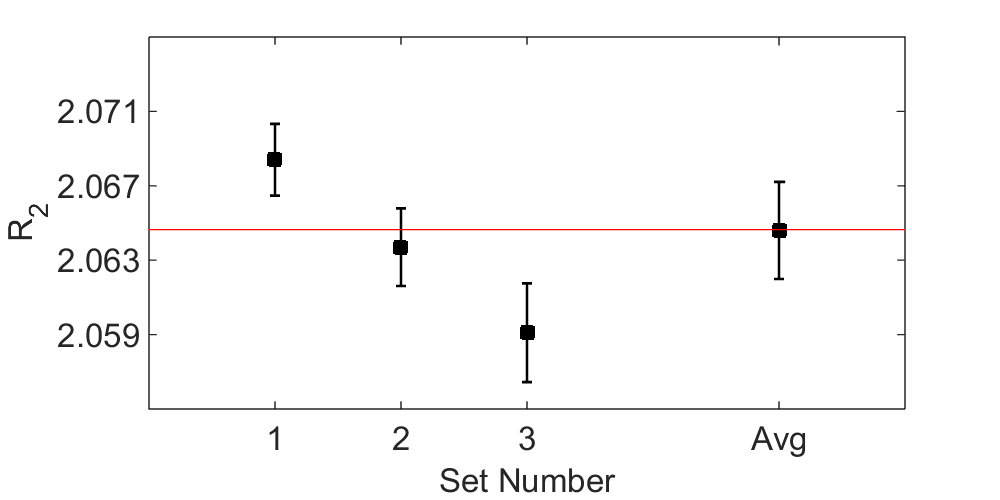}\\
  \caption{ 
	  Plot of the three sets of $R_2$ results and the weighted mean.  We have expanded the error bars of the weighted mean by $\sqrt{\chi_r^2}= 2.0$ to account for the variation among the individual results.} 
	  \label{fig:R2}
\end{figure}

\begin{figure}
  \includegraphics[width=\columnwidth]{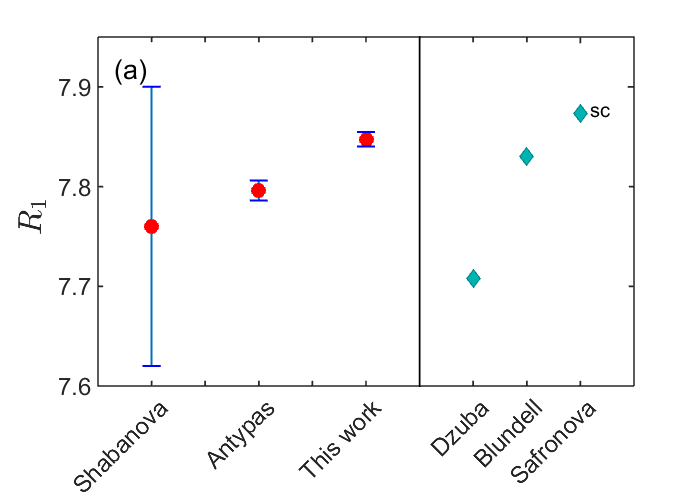}\\
  \includegraphics[width=\columnwidth]{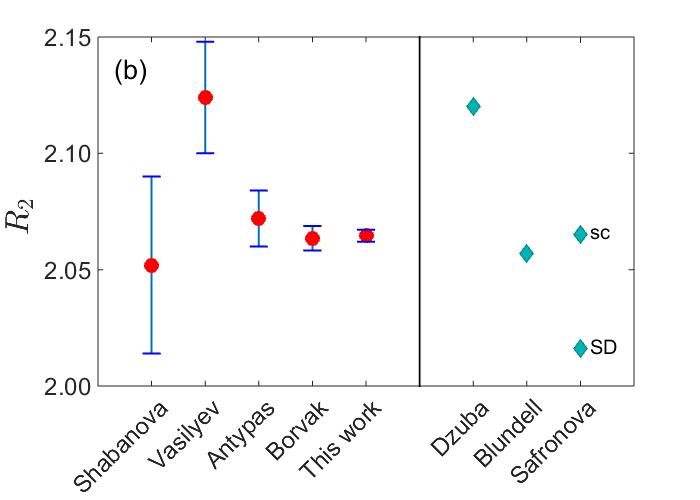}\\
	  \caption{Plots showing comparisons of various experimental and theoretical determinations of (a) $R_1 \equiv \langle 6s_{1/2} || r ||   6p_{1/2}\rangle / \langle 6s_{1/2} || r ||  7p_{3/2}\rangle$ and (b) $R_2 \equiv \langle 6s_{1/2}  || r ||  $ $ 7p_{3/2}\rangle / \langle 6s_{1/2} || r ||  7p_{1/2}\rangle$ ratios. See Table \ref{table:ResultComparison} for references to these data.  For Safronova {\it et al.} (Refs.~\cite{SafronovaJD99,SafronovaSC16}), we plot both the scaled (sc) and single-double (SD) values for $R_2$.
	  } 
	  \label{fig:R1R2}
\end{figure}

After adding the uncertainties described in the previous section, the final result for $R_1$ is $R_1 = 7.8474 \: (72)$.  
For $R_2$, after applying the corrections and uncertainties described in the previous section to the three individual data sets, we arrive at the results shown in Table \ref{table:R2results} and plotted in Fig.~\ref{fig:R2}. The weighted average of these results is $R_2 = 2.0646 \: (26)$. 
We compare these results for $R_1$ and $R_2$ with a number of prior experimental and theoretical results in Table \ref{table:ResultComparison}, and illustrate these in Fig.~\ref{fig:R1R2}.  
We note that the previous results for $R_1$ and $R_2$ by Shabanova {\it et al.} \cite{shabanovaMK1979} (who reported oscillator strengths, which we converted to matrix elements) and $R_2$ by Borv\'{a}k \cite{Borvak14} are in reasonable agreement, to within their error bars, of our results, which are of higher precision.  (We derived the ratio value for Borv\'{a}k using $(R_2)^2 = (21/22) \times 4.461 \: (23)$ from the table on page 126, where the additional factor of $21/22$ comes from combinations of Clebsch-Gordan coefficients.)  $R_1$ from Antypas \cite{antypas7p2013} disagrees with our new results, which we consider to be more reliable due to the use of the Voigt profile and the addition of the faster photodiode amplifier that we discussed earlier.  We derived the value of $R_2$ for Vasilyev {\it et al.}~\cite{VasilyevSSB02} from their matrix elements, whose measurement was dependent on precise knowledge of the vapor cell path length and atomic density in the vapor cell, unlike our method. The scaled theoretical results of Refs.~\cite{BlundellSJ92, SafronovaJD99, SafronovaSC16} appear to be in good agreement with our new results as well.

We use $R_1 = 7.8474 \: (72)$ to determine the matrix element for the $6s \ ^2S_{1/2} \rightarrow 7p \ ^2P_{3/2}$ transition using 
\begin{equation}
  \langle 6s_{1/2} || r ||  7p_{3/2}\rangle = \frac{\langle 6s_{1/2} || r ||  6p_{1/2} \rangle}{R_1} 
\end{equation} 
and $\langle 6s_{1/2}  || r || 6p_{1/2}\rangle = 4.5057 \: (16) \ a_0$, the weighted average of the transition matrix element for the D$_1$ line from Refs.~\cite{YoungHSPTWL94,RafacT98,RafacTLB99,DereviankoP02a,AminiG03,BouloufaCD07,ZhangMWWXJ13,PattersonSEGBSK15,GregoireHHTC15,TannerLRSKBYK92,SellPEBSK11}.
Our result is 
\begin{equation}
\langle 6s_{1/2} || r ||  7p_{3/2}\rangle = 0.57417 \: (57) \ a_0.
\end{equation}

\begin{table*}[t]
  \caption{Experimental and theoretical results for the reduced dipole matrix elements of the cesium $6s\,^2S_{1/2} \rightarrow 7p\,^2P_{J}$ transitions. The matrix elements are given as factors of $a_0$.  For Ref.~\cite{SafronovaSC16}, we list both the single double (-SD) and scaled (-sc) values.}
  \begin{tabular}{|l|l|l|l|l|}
    \hline
      \multicolumn{1}{|c|} {Group}  & \multicolumn{1}{|c|}{$R_1$} & \multicolumn{1}{|c|}{$R_2$} & $\langle 6s_{1/2}||r|| 7p_{3/2} \rangle$ & $\langle 6s_{1/2}||r|| 7p_{1/2} \rangle$  \\ \hline 
  {\underline{\emph{Experimental}}} 		& & & &		 \\
   Shabanova {\it et al.}, 1979~\cite{shabanovaMK1979} 	& \hspace{0.1in} 7.76 (14)       & \hspace{0.1in} 2.052 (38) & \hspace{0.1in} 0.583 (10)   & \hspace{0.1in} 0.2841 (21)  \\
   Vasilyev {\it et al.}, 2002~\cite{VasilyevSSB02} 		&  & \hspace{0.1in} 2.124 (24) &  \hspace{0.1in} 0.5856 (50)  & \hspace{0.1in} 0.2757 (20)  \\
   Antypas and Elliott, 2013~\cite{antypas7p2013}    	& \hspace{0.1in} 7.796 (41)      & \hspace{0.1in} 2.072 (12) &  \hspace{0.1in} 0.5780 (7)   & \hspace{0.1in} 0.2789 (16)  \\
   Borv\'{a}k, 2014~\cite{Borvak14}			&  & \hspace{0.1in} 2.0635 (53)~~ &  \hspace{0.1in} 0.5759 (30)  & \hspace{0.1in} 0.2743 (29)  \\
   This work	                  			& \hspace{0.1in} 7.8474 (72)~~ & \hspace{0.1in} 2.0646 (26) &  \hspace{0.1in} 0.57417 (57) & \hspace{0.1in} 0.27810 (45) \\
   & & & & \\
  \multicolumn{1}{|l|}{\underline{\emph{Theoretical}}} 	& & & &  \\
    Dzuba {\it et al.}, 1989~\cite{DzubaFKS89}			& \hspace{0.1in} 7.708 & \hspace{0.1in} 2.12 & \hspace{0.1in} 0.583 	& \hspace{0.1in} 0.275 \\
    Blundell {\it et al.}, 1992~\cite{BlundellSJ92}	    & \hspace{0.1in} 7.83 & \hspace{0.1in} 2.057 & \hspace{0.1in} 0.576 	& \hspace{0.1in} 0.280 \\
    Safronova {\it et al.}, 1999~\cite{SafronovaJD99}	& \hspace{0.1in} 7.873 & \hspace{0.1in} 2.065 & \hspace{0.1in} 0.576 	& \hspace{0.1in} 0.279 \\
    Derevianko, 2000~\cite{Derevianko00}	&  &  &  & \hspace{0.1in} 0.281 \\
    Porsev {\it et al.}, 2010~\cite{PorsevBD10} 			&  &  &  & \hspace{0.1in} 0.2769 \\
    Safronova {\it et al.}-SD, 2016~\cite{SafronovaSC16}	& \hspace{0.1in} 7.452 & \hspace{0.1in} 2.016 & \hspace{0.1in} 0.601     & \hspace{0.1in} 0.298 \\
    Safronova {\it et al.}-sc, 2016~\cite{SafronovaSC16} & \hspace{0.1in} 7.873 & \hspace{0.1in} 2.065 & \hspace{0.1in} 0.576     & \hspace{0.1in} 0.279 \\
   \hline
  \end{tabular}
  \label{table:ResultComparison}
\end{table*}

\begin{figure}
    \includegraphics[width=\columnwidth]{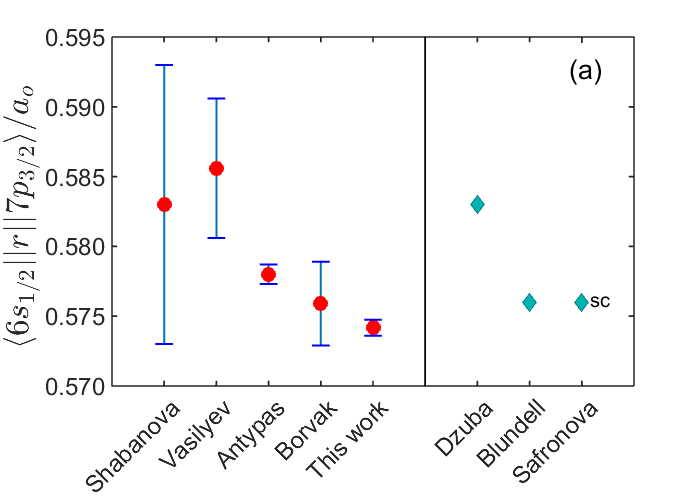}\\
    \includegraphics[width=\columnwidth]{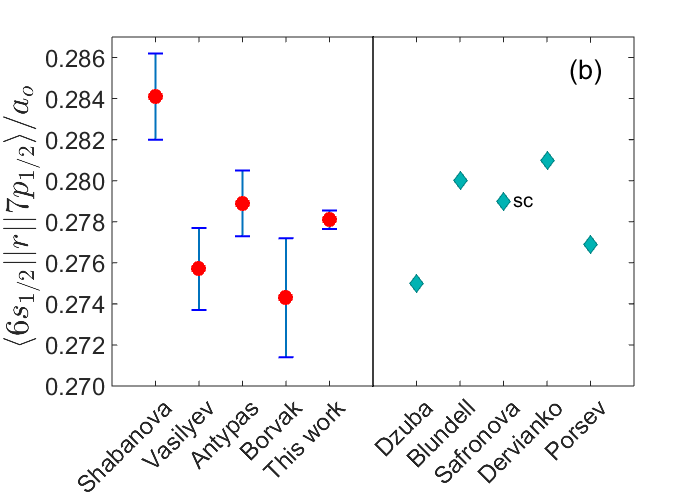}\\
	  \caption{ Plots of the experimental and theoretical values of (a) $\langle 6s_{1/2}||r|| 7p_{3/2} \rangle$ and (b) $\langle 6s_{1/2}||r|| 7p_{1/2} \rangle$, as listed in Table ~\ref{table:ResultComparison}. Experimental values are on the left, while theoretical values are shown on the right.  See Table \ref{table:ResultComparison} for references to these data.  For Safronova {\it et al.} (Refs.~\cite{SafronovaJD99, SafronovaSC16}), we have plotted only the scaled (sc) values.} 
	  \label{fig:matrix1232}
\end{figure}

We combine $R_2 = 2.0646 \: (26)$ in Eq.~(\ref{eq:ratioR2}) and our new determination of $\langle 6s_{1/2}  || r || 7p_{3/2}\rangle$ to obtain
\begin{equation}
\langle  6s_{1/2}|| r ||7p_{1/2} \rangle = 0.27810 \: (45)~a_0.
\end{equation}

We present a summary of past experimental and theoretical results of these dipole matrix elements in Table~\ref{table:ResultComparison}. We have also plotted the results for $\langle 6s_{1/2}  || r || 7p_{3/2}\rangle$ and $\langle 6s_{1/2}  || r || 7p_{1/2}\rangle$ in Figs.~\ref{fig:matrix1232}(a) and (b), respectively.  
For Shabanova {\it et al.}~\cite{shabanovaMK1979} and Borv\'{a}k \cite{Borvak14}, our result is within their uncertainties for $\langle 6s_{1/2}|| r ||  7p_{3/2} \rangle$, but in poorer agreement with the $\langle 6s_{1/2}|| r ||  7p_{1/2} \rangle$ value.  The matrix element values from Borv\'{a}k come from a direct determination, separate from the ratio measurement discussed above.

For comparison with theory, our result is within the distribution spanned by the $\langle 6s_{1/2}||r|| 7p_{1/2} \rangle$ values. In particular, our result is in the middle of the two closest theory values from Refs.~\cite{SafronovaJD99, PorsevBD10}. For $\langle  6s_{1/2} || r || 7p_{3/2} \rangle$, our value is below all of the theoretical results, but within 0.3\% of Blundell {\it et al.}~\cite{BlundellSJ92} and scaled values of Refs.~\cite{SafronovaJD99, SafronovaSC16}. 
In Table~\ref{table:ResultComparison}, we list two values, calculated results using different theoretical methods, from the Supplemental Material of Ref.~\cite{SafronovaSC16}. 
The authors of \cite{SafronovaSC16} recommend the single-double (SD) all-order approach values, which we have listed as `Safronova {\it et al.}-SD'. In Ref.~\cite{SafronovaJD99}, the authors note that scaling improved agreement of theoretical determinations of the $\langle 6s_{1/2}  || r || 7p_{j}\rangle$ matrix elements with experiment, and recommend using the scaled values. We also observe that scaling improves the agreement of theory with the current measurements, and have included those values from \cite{SafronovaSC16} as `Safronova {\it et al.}-sc'.
In comparison with Refs.~\cite{SafronovaJD99, SafronovaSC16}, our measurements of both matrix elements are much closer to the scaled theory values than the SD theory values.

\section{Conclusion}\label{sec:Conclusion}

In conclusion, we present measurements of the ratio of the dipole matrix elements of the cesium $6s\,^2S_{1/2} \rightarrow 6p\,^2P_{1/2}$ and $6s\,^2S_{1/2} \rightarrow 7p\,^2P_{3/2}$ transitions and the ratio of $6s\,^2S_{1/2} \rightarrow 7p\,^2P_{1/2}$ and $6s\,^2S_{1/2} \rightarrow 7p\,^2P_{3/2}$ transitions. We used a ratio measurement of the two transitions to eliminate the need for precise knowledge of the path length of the laser within the vapor cell, or of the density of cesium, which helped to eliminate potential systematic errors. From these measurements, we calculate new, higher precision results of the dipole matrix elements of cesium with precision $\le 0.16 \%$. With our new knowledge of the dipole matrix elements, we are poised to be able to evaluate the scalar and vector polarizabilities of the cesium $6s\,^2S_{1/2} \rightarrow 7s\,^2s_{1/2}$ transition. A new value of the vector polarizability has implications on the interpretation of cesium parity nonconservation measurements, and will allow a new determination of the weak charge in cesium.

This material is based upon work supported by the National Science Foundation under Grant Number PHY-1607603 and PHY-1460899.  Useful conversations with Dan Leaird are also gratefully acknowledged.
\bibliography{biblio}

\end{document}